\documentclass[prb,twocolumn,aps,superscriptaddress,floatfix]{revtex4}
\usepackage{amsmath}
\usepackage{amssymb}
\usepackage{graphicx}
\usepackage{color}
\usepackage[svgnames]{xcolor}
\usepackage[normalem]{ulem}
\usepackage{cancel}
\usepackage{hyperref}
\usepackage{verbatim}

\begin{document}

\title{Umklapp-driven first-order transition into a charge-density wave
phase}

\author{A.V. Rozhkov}
\affiliation{Institute for Theoretical and Applied Electrodynamics, Russian
Academy of Sciences, 125412 Moscow, Russia}

\begin{abstract}
Using phenomenological Landau free energy, we systematically analyze
influence of so-called umklapp contributions on the phase diagram of a
solid hosting a commensurate charge-density wave phase. The umklapp terms
may change the transition into the ordered state from continuous
(second-order) to discontinuous (first-order). Additional elements, such as
critical and tricritical points, emerge on the phase diagram under suitable
conditions. The proposed mechanism is generalized for the case of
nearly-commensurate charge-density wave, as well as spin-density wave
orders. Several charge-density-wave-hosting alloys are used as
experimentally available examples illustrating the formalism.
\end{abstract}

\date{\today}

\maketitle

\section{Introduction}
\label{intro}

Canonical theory of the charge-density wave (CDW) thermodynamic phase
(e.g., 
Ref.~\onlinecite{gruner_book})
concludes that the transition from a disordered phase into a CDW state is
continuous (second-order). This basic theoretical expectation is indeed
confirmed by numerous experiments. For example, 
Fig.~5
in review
paper~\onlinecite{gruner_review_dynamics1988}
demonstrates continuous decay of the CDW order parameter to zero as a
function of increasing temperature, for three different CDW-hosting alloys
[NbSe$_3$, (TaSe$_4$)$_2$I, and K$_{0.3}$MoO$_3$].
As a more recently published instance of the same behavior, we can mention
Fig.~3 in
Ref.~\onlinecite{cdw2017cont},
where order-parameter-versus-temperature data for
TiSe$_2$
are presented.
Reference~\onlinecite{cdw_fluct2008exper}
examined the transition type for
TbTe$_3$.
Gradual decay of the order parameter, absence of hysteresis, and critical
fluctuations all point to the continuous transition in the latter compound.

Yet for many crystals a CDW phase is separated from a disordered state by a
discontinuous (first-order) transition. Namely, in
IrTe$_2$
the formation of a commensurate CDW (CCDW) at temperature
$\sim 280$\,K
is accompanied by pronounced hysteresis inside heating-cooling
cycle~\cite{IrTe2021hysteresis_exper, IrTe2014first_order_exper,
IrTe2015first_exper,IrTe2kim2015dft_arpes, Ivashko2017},
the signature of a first-order transition. Another alloy demonstrating 
the first-order transition between a CCDW and a disordered state is
Lu$_5$Ir$_4$Si$_{10}$,
with the CDW
transition~\cite{Lu5Ir4Si10first_orderCDW1999exper}
at 83\,K.

In the interval between 130\,K and 150\,K a discontinuous transition into a
nearly-commensurate CDW (NC-CDW) phase was 
reported~\cite{Er2Ir3Si5_first_order_CDW2020exper}
for
Er$_2$Ir$_3$Si$_5$.
Similarly,
Lu$_2$Ir$_3$Si$_5$
enters NC-CDW phase through a first-order
transition~\cite{Lu2Ir3Si52005hyster_exper,
Lu2Ir3Si5_cdw2006hysteresis_exper, Lu2Ir3Si5_first_orderCDW2015exper},
with transition temperature
$\sim 200$\,K.

Various theoretical
mechanisms~\cite{young1974role,
TaSe2McMillan1975theory_umklapp,
fine2008phase_separation,
sboychakov2013electronic,
tritelluride2019hysteresis_compet,
q1d2023landau_1st_order, q1d2023first_order,
kagan2021electronic},
each with its own applicability range, are employed to explain the origins
of discontinuous transition into a density-wave state. The most relevant
for us here is a phenomenological approach pioneered in
Ref.~\onlinecite{TaSe2McMillan1975theory_umklapp}
to describe a CDW phase in
TaSe$_2$.

The Landau-Ginzburg functional, specifically designed in the latter
reference to account for peculiar features a CDW order, serves as a
cornerstone of modern CDW modeling. Important concepts, such as ``lock-in''
NC-CDW-to-CCDW transition and order parameter ``dislocations'', were
discussed. Investigation in
Ref.~\onlinecite{TaSe2McMillan1975theory_umklapp}
revealed that the so-called ``umklapp'' contributions to the Landau free
energy can transform the continuous transition into a discontinuous one.

Here, we aim to explore systematically the influence of the umklapp terms,
especially those of higher order, on the phase diagram of a CDW-hosting
material. Constructing a suitable Landau free energy, we study phase
transitions into commensurate and nearly-commensurate CDW states. We argue
that in various rather general situations such transitions may become
discontinuous. Moreover, for some conditions our theory predicts that a
material can demonstrate two-transition sequence: a continuous
normal-to-CCDW transition is followed by a first-order CCDW-CCDW
transition, the latter connecting the states that differ only by order
parameter magnitudes. (Note that a ``lock-in'' commensurate-incommensurate
transition~\cite{TaSe2McMillan1975theory_umklapp}
is a phenomenon of a distinctively dissimilar type, which will not be
investigated below.)

As for NC-CDW, we show that, if non-CDW lattice distortions are
incorporated into the model, suitably constructed umklapp terms become
symmetry-allowed. When the lattice distortions fields are eliminated from
the free energy, the resultant effective model is equivalent to CCDW Landau
free energy, and a first-order transition can be recovered. We speculate
that this scenario may explain a first-order transition within NC-CDW phase
of
EuTe$_4$
reported in
Refs.~\onlinecite{EuTe4hyster2019exper, EuTe4hysteresis2022exper,
lv2025EuTe4}.

Our paper is organized as follows.
Section~\ref{sec::general_matters} 
is dedicated to formulation of a Landau free energy function valid for
unidirectional incommensurate CDW. Various models of CCDW are introduced
and analyzed in
Sec.~\ref{sec::ccdw}.
The case of NC-CDW is presented in
Sec.~\ref{sec::near_commensurate}.
Our results are discussed in
Sec.~\ref{sec::discussion}.
Technically involved calculations secondary to the main presentation are
relegated to Appendices.

\section{General considerations}
\label{sec::general_matters} 

It is common to describe transition into a CDW state within the framework
of the Landau free energy 
\begin{eqnarray}
\label{eq::Landau_U1}
F_0 (\rho_{\rm cdw})
=
\frac{a}{2} |\rho_{\rm cdw}|^2 + \frac{b}{4} |\rho_{\rm cdw}|^4,
\end{eqnarray}
where the coefficients $a$ and $b$ satisfy the well-known conditions
$b>0$
and
$a (T) = \alpha (T - T_{\rm CDW})$.
Here $T$ is temperature, 
$T_{\rm CDW}$
is the CDW transition temperature, and coefficient $\alpha$ is positive.

As for the complex order parameter
$\rho_{\rm cdw}
=
|\rho_{\rm cdw}| e^{i \varphi}$,
it represents periodic charge-density modulation. In many situations it is
conveniently approximated by a single harmonic term
\begin{eqnarray}
\rho( {\bf R} )
\approx
\rho_{\rm cdw} e^{i {\bf k} \cdot {\bf R}}
+ 
{\rm c.c.}
=
2 |\rho_{\rm cdw}| \cos ( {\bf k} \cdot {\bf R} + \varphi) ,
\end{eqnarray} 
where the wave vector
${\bf k}$
characterizes CDW spatial periodicity. Following the standard prescription,
one minimizes
$F_0$
over
$\rho_{\rm cdw}$
to derive
\begin{eqnarray}
|\rho_{\rm cdw}|
= \theta( T_{\rm CDW} - T) \sqrt{\frac{\alpha ( T_{\rm CDW} - T) }{b}},
\end{eqnarray} 
where
$\theta(x)$
is the Heaviside step-function. This formula explicitly demonstrates that
the order parameter strength
$|\rho_{\rm cdw}|$
is a continuous function of $T$, a hallmark of the second-order transition. 

Unlike the absolute value
$|\rho_{\rm cdw}|$,
the order parameter phase $\varphi$ remains undetermined, which is a
manifestation of the U(1)-symmetry of
$F_0$:
the change
\begin{eqnarray} 
\varphi \rightarrow \varphi + \delta \varphi
\end{eqnarray} 
keeps
$F_0$
the same. Physically, this is a consequence of CDW free energy being
invariant under arbitrary uniform translation 
\begin{eqnarray}
\label{eq::translation_cont}
{\bf R} \rightarrow {\bf R} + {\bf t},
\quad
{\bf t} \in \mathbb{R}^3.
\end{eqnarray} 
It is easy to check that, for a given
${\bf t}$,
the phase change is
$\delta \varphi = ({\bf k} \cdot {\bf t})$,
or, equivalently, the order parameter transforms according to
\begin{eqnarray}
\label{eq::rho_transform}
\rho_{\rm cdw} \rightarrow \rho_{\rm cdw} e^{i ({\bf k} \cdot {\bf t})}
\end{eqnarray} 
under the translation
${\bf t}$.

\section{Commensurate CDW}
\label{sec::ccdw}

\subsection{Landau free energy with `umklapp' contribution}

The invariance of the CDW Landau free energy relative to arbitrary
translations~(\ref{eq::translation_cont})
is, by itself, a very excessive constraint on the model: in any crystal
the translation group must be limited to lattice translations only, that
is, instead of
${\bf t} \in \mathbb{R}^3$,
the allowed 
${\bf t}$'s
are
\begin{eqnarray}
\label{eq::lattice_translation}
{\bf t} = m_1 {\bf a}_1 + m_2 {\bf a}_2 + m_3 {\bf a}_3,
\end{eqnarray} 
where 
$m_i$
are integers, and
${\bf a}_i$
are elementary lattice vectors.

The reduction of the invariance group implies that additional terms may be
introduced into the Landau free energy. Below we explicitly construct these
terms for commensurate CDW order. 

By definition, a commensurate CDW satisfies the following conditions
\begin{eqnarray}
({\bf k} \cdot {\bf a}_i ) = \frac{2 \pi p_i}{q_i},
\quad
i = 1,2,3,
\end{eqnarray}
where integer 
$p_i$
is co-prime with
$q_i$
for all $i$. Formally, of course, any measured 
${\bf k}$
can be described in this manner, with arbitrary large
$q_i$'s.
However, for practical matters, a wave vector is considered to be
commensurate only when all three
$q_i$'s
are not too large.

For these three 
$q_i$
we introduce their least common multiple
$n = {\rm lcm} (q_1, q_2, q_3)$,
referred below to as commensuration degree. Then the monomial
$\rho_{\rm cdw}^n$
is invariant under arbitrary lattice translations. To prove this claim, we
start with
Eq.~(\ref{eq::rho_transform})
and write
$\rho_{\rm cdw}^n
\rightarrow
\rho_{\rm cdw}^n e^{i ({\bf k} \cdot {\bf t}) n}$,
where
\begin{eqnarray} 
\label{eq::commensuration_general}
({\bf k} \cdot {\bf t}) n
=
2\pi \sum_i m_i p_i \frac{n}{q_i}.
\end{eqnarray} 
Since $n$ is a multiple of 
$q_i$
for any $i$, one establishes that
$({\bf k} \cdot {\bf t}) n = 2\pi N$,
where $N$ is an integer. Thus, 
$\rho_{\rm cdw}^n$
is invariant for any 
${\bf t}$
described by
Eq.~(\ref{eq::lattice_translation})

Note that, while $n$ is formally introduced as the least common multiple of
the three denominators, in many realistic situations, however, no
significant number-theoretical calculations are required, as the
commensuration degree is quite obvious from the data. For example, if 
$(p_1/q_1, p_2/q_2, p_3/q_3) = (1/5, 0, 1/5)$,
as in
Ref.~\onlinecite{Ivashko2017},
then
$n=5$.

Since
$\rho_{\rm cdw}^n$
is invariant, we conclude that, for any complex number
$c_n = |c_n| e^{i \gamma}$,
the $n$th degree `umklapp' contribution
\begin{eqnarray}
\label{eq::F_clock}
F_{\rm u}^{(n)} = - \frac{c_n}{2n} \rho_{\rm cdw}^n + {\rm c.c.}
=
-\frac{|c_n|}{n} |\rho_{\rm cdw}|^n \cos ( n \varphi + \gamma)
\end{eqnarray} 
is explicitly real and invariant under lattice translations (similar
expressions for the umklapp term, in a variety of settings, can be
found~\cite{pnas2012umklapp, jaramillo2009breakdown, LEE1974703,
young1974role}
in published literature).

Thus, the free energy
\begin{eqnarray}
\label{eq::landau_free_commens}
F^{(n)} = F_0 + F_{\rm u}^{(n)}
\end{eqnarray}
can be used as a model for a commensurate CDW state. Note that inclusion of
$F_{\rm u}^{(n)}$
shrinks the symmetry group of the Landau energy from U(1) to
$Z_n$:
function 
$F^{(n)}$
is no longer invariant under an arbitrary phase shift, only discrete
shifts
\begin{eqnarray}
\varphi \rightarrow \varphi + \frac{2\pi m}{n}, \quad m \in \mathbb{Z},
\end{eqnarray} 
do not change the free energy. 

Let us now search for minima of
$F^{(n)}$.
Minimization with respect to $\varphi$ is very simple. As this variable
enters
$F_{\rm u}^{(n)}$
term only, it is easy to demonstrate that the Landau energy is the lowest
when
$\varphi = \varphi^*_m$,
where
\begin{eqnarray}
\label{eq::clock_minima}
\varphi^*_m = \frac{2\pi m}{n} - \frac{\gamma}{n},
\quad
m = 0, 1, \ldots, n-1.
\end{eqnarray} 
We see $n$ minima evenly distributed over a unit circle. This arrangement
resembles a clock dial, thus a common name for a model of this kind is the
$n$-state clock model. Another frequently used designation is the
$Z_n$
model, a reference to the invariance group of
$F^{(n)}$.

\begin{figure}[t!]
\centering
\includegraphics[width=0.99\columnwidth]{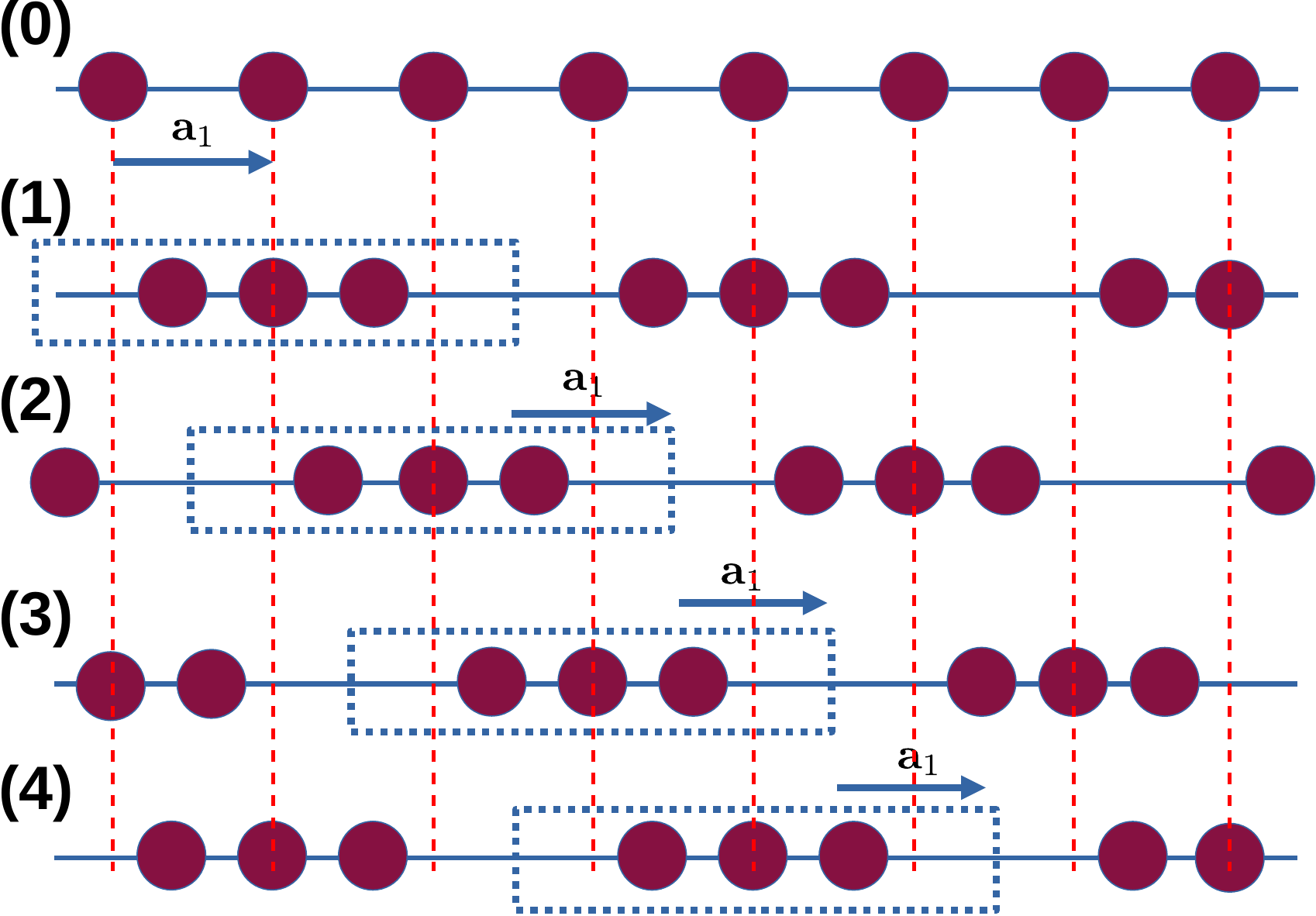}
\caption{Non-equivalent lattice configurations for a commensurate CDW for
$n=3$.
Five panels schematically represent the same lattice with and without CCDW
distortions. Primitive vector
${\bf a}_1$
of the unperturbed lattice (panel~0) is drawn as a (blue) arrow. The
directions orthogonal to
${\bf a}_1$
are not depicted.
Panels from~1 to~4 show the lattice distorted by the CDW. The CDW unit cell
(dotted-line rectangle) grows three-fold relative to the unit cell of the
pristine lattice. Vertical (red) dashed lines mark undistorted atoms
positions. Starting from the CDW configuration in panel~1, one can generate
two more structures (panels~2 and~3) by executing two consecutive
translations on 
${\bf a}_1$.
The third translation does not produce a new structure, instead the
distortion shown in panel~1 is recovered, as indeed panel~4 demonstrates.
Each configuration in panels~1, 2, and~3 represents one of three minima of
the
$Z_3$ model.
\label{fig::cdw3}
}
\end{figure}

Every 
$\varphi^*_m$
in
Eq.~(\ref{eq::clock_minima})
represents a particular localization of CDW distortions relative to the
underlying lattice, see
Fig.~\ref{fig::cdw3}.
There are $n$ such localizations, all of them are degenerate. Speaking
heuristically, one can say that a CDW with $n$th degree commensuration is
always pinned by the lattice to one of $n$ possible minima, as illustrated
by
Fig.~\ref{fig::cdw3}.
This pinning decreases the symmetry of the Landau free energy from U(1) to
$Z_n$.

At any of these minima
the cosine in
Eq.~(\ref{eq::F_clock})
is equal to unity. Thus, the Landau free energy can be re-written as a
function of a single non-negative variable
$|\rho_{\rm cdw}|$
\begin{eqnarray}
\label{eq::Fn_def}
\Tilde F^{(n)}
=
\frac{a}{2} |\rho_{\rm cdw}|^2 + \frac{b}{4} |\rho_{\rm cdw}|^4
- \frac{|c_n|}{n} |\rho_{\rm cdw}|^n
+ \ldots,
\end{eqnarray} 
where ellipses stand for higher-order terms that might be required to
maintain stability of the free energy. The tilde over $F$ implies that
this free energy does not depend on $\varphi$.

For
$n=2$
the contribution proportional to
$|c_2|$
acts to renormalize $a$, effectively increasing the transition temperature.
The transition remains continuous for all 
$|c_2|$.
If 
$n>2$,
the contribution coming from the umklapp term
$F^{(n)}_{\rm u}$
may qualitatively alter the behavior of the system near the transition
point, as discussed below.

\subsection{$Z_3$ model of CDW}
\label{subsec::n3}

The value
$n=3$
represents the CCDW phase whose unit cell is three times larger
than the unit cell of the underlying lattice. (This type of order is
schematically shown in
Fig.~\ref{fig::cdw3}.)
Specializing 
Eq.~(\ref{eq::Fn_def})
for
$n=3$,
one can express the free energy 
$\Tilde F^{(3)}$
as
\begin{eqnarray}
\Tilde F^{(3)} (|\rho_{\rm cdw}|)
=
\frac{a}{2} |\rho_{\rm cdw}|^2 - \frac{|c_3|}{3} |\rho_{\rm cdw}|^3
+ \frac{b}{4} |\rho_{\rm cdw}|^4.
\end{eqnarray} 
We see that this free energy is stable in the sense that, for large 
$|\rho_{\rm cdw}|$,
function
$\Tilde F^{(3)}$
grows, which guarantees that an equilibrium value of the order parameter
is bounded.
\begin{figure}[t!]
\centering
\includegraphics[width=0.99\columnwidth]{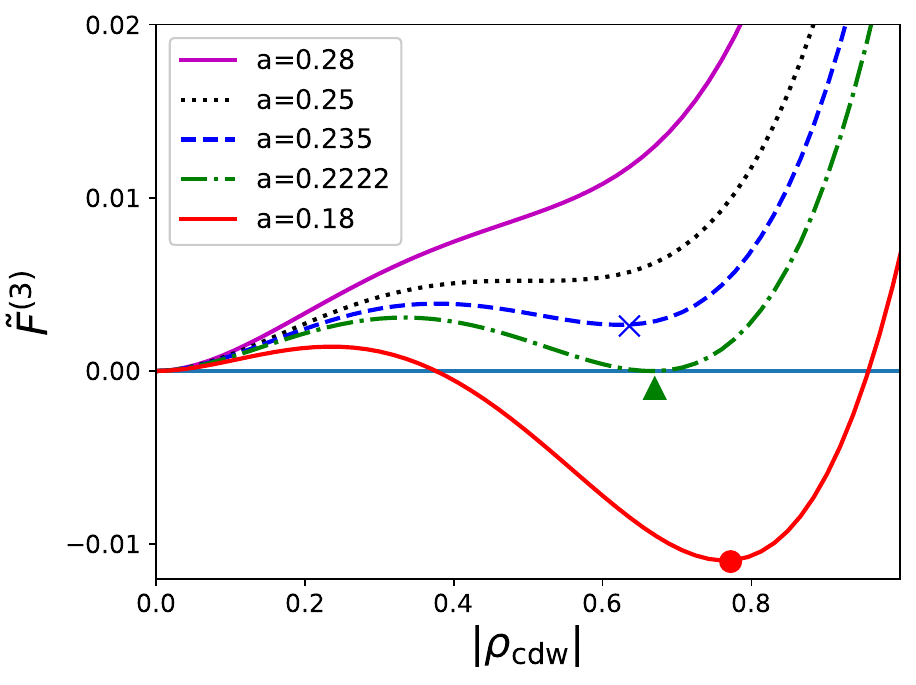}
\caption{Landau free energy 
$\Tilde F^{(3)}$
as a function of the order parameter 
$|\rho_{\rm cdw}|$,
for various temperatures (various values of $a$, see legend). The graphs
here are plotted for
$b=1$
and
$|c_3|=1$,
all units are arbitrary. The solid (magenta) curve with a single minimum at
$|\rho_{\rm cdw}| = 0$
represents the system in the high-temperature disordered phase. The dashed
(blue) curve with a non-trivial minimum (marked by the cross) shows the
formation of the metastable CDW state at lower temperature (lower $a$). The
dotted (black) curve separates the curves with and without a non-zero
minimum. This separatrix is realized when
$a(T)=|c_3|^2/(4b)$.
If the trivial and non-trivial minima have identical free energies, which
is the case of the dash-dotted (green) curve, the first-order phase
transition occurs. Discontinuity of the order parameter at the transition
is marked by (green) triangle. Solid (red) curve correspond to ordered
phase, with the circle marking the stable value of
$|\rho_{\rm cdw}|$.
The disordered phase
$|\rho_{\rm cdw}|=0$
is a metastable minimum on this curve. 
\label{fig::n3} 
}
\end{figure}

For
$a>0$
this free energy has a
$|\rho_{\rm cdw}|=0$
minimum that represents (meta)stable disordered state. Additionally, for
$a < |c_3|^2/(4b)$
there is a minimum of
$\Tilde{F}^{(3)}$
at
\begin{eqnarray}
\label{eq::min_n3}
|\rho_{\rm cdw}|
= 
\frac{1}{2b} \left( |c_3| + \sqrt{|c_3|^2 - 4ab} \right),
\end{eqnarray} 
see
Fig.~\ref{fig::n3}.
It is easy to check that 
Eq.~(\ref{eq::min_n3})
describes the global minimum of 
$\Tilde F^{(3)}$
when
$a < 2|c_3|^2/(9b)$.

By exploiting the commonly assumed linearization
\begin{eqnarray} 
\label{eq::a_linear}
a = a(T) \approx \alpha (T-T_*),
\quad
\alpha > 0,
\end{eqnarray} 
where 
$T_*$
is the temperature for which
$a(T)$
passes through zero, the CDW transition temperature can be expressed as
\begin{eqnarray}
T_{\rm CDW} = T_* + \frac{2 |c_3|^2}{9 \alpha b} > T_*.
\end{eqnarray} 
We see that
$T_*$
by itself does not have any special meaning. However, in the limit
$|c_3| \rightarrow 0$
the transition temperature
$T_{\rm CDW}$
approaches
$T_*$.

At the transition, the coefficient $a$ is not zero, but rather 
$a = 2|c_3|^2/(9b)$.
Substituting this value in
Eq.~(\ref{eq::min_n3}),
one finds that
$|\rho_{\rm cdw}|$
jumps from 0 to
$2|c_3|/(3b)$,
see also
Fig.~\ref{fig::n3}.
Thus, we demonstrate that, at finite
$|c_3|$,
the transition is discontinuous. On the other hand, one must remember that
at small
$|c_3|$
the transition is formally indeed first-order, yet, in this regime, the
discontinuity of order parameter becomes weak, and difficult to detect.
This observation remains relevant for other signatures of first-order
transition.

Concluding the discussion of the
$Z_3$-type
models, we would like to note that, in the context of the CDW theory, cubic
contributions to the Landau free energy are known for quite some time, at
least since
Ref.~\onlinecite{TaSe2McMillan1975theory_umklapp},
where a CDW state in
TaSe$_2$
was studied. 

Note, however, that despite mathematical equivalence, the physical origins
of the cubic term may differ, depending on specific details. In our
derivation, we relied on 
condition~(\ref{eq::commensuration_general})
at
$n=3$
that demands of
$3 {\bf k}$
to be equal to a reciprocal lattice vector. Alternatively, when the three
wave vectors of three co-existing CDW order parameters sum up to zero, a
cubic term may emerge independently of
requirement~(\ref{eq::commensuration_general}).
Reference~\onlinecite{TaSe2McMillan1975theory_umklapp}
discussed both of these mechanisms.

\subsection{$Z_4$ model of CDW}
\label{subsec::n4}

The
$n=4$
phase diagram differs qualitatively from the
$n=3$
situation. The 
$n=4$
Landau free energy reads
\begin{eqnarray}
\Tilde F^{(4)}
=
\frac{a}{2} |\rho_{\rm cdw}|^2 + \frac{\Tilde b}{4} |\rho_{\rm cdw}|^4
+ \frac{d}{6} |\rho_{\rm cdw}|^6,
\end{eqnarray} 
where
$\Tilde b = b - |c_4|$.
In other words,
$|c_4|$
effectively renormalizes $b$. Since 
$\Tilde b$
can be either positive, or negative, depending on the relation between $b$
and
$|c_4|$,
we retained here the sixth-order term to prevent uncontrollable growth of
the order parameter at 
$\Tilde b < 0$.

When
$|c_4| < b$,
the free energy 
$\Tilde F^{(4)}$
describes second-order transition that occurs at
$a=0$.
If
linearization~(\ref{eq::a_linear})
is assumed, then the transition temperature coincides with
$T_*$.

At negative
$\Tilde b$,
the transition into the CDW phase becomes first-order. (Qualitatively, the
behavior of
$\tilde F^{(4)}$
in this regime is very similar to the graphs of
$\tilde F^{(3)}$
in
Fig.~\ref{fig::n3}.)
For
$0<4ad < \Tilde b^2$
the free energy has three extrema: one at zero, and two more at
\begin{eqnarray}
|\rho_{\rm cdw}|
=
\sqrt{
	\frac{1}{2d} \left( |\Tilde b| \pm \sqrt{\Tilde b^2 - 4ad} \right)
}.
\end{eqnarray} 
The minimum (maximum) corresponds to the plus (minus) sign in this formula.
The transition into the ordered state takes place when the free energy at
the non-trivial minimum becomes zero, which is the free energy at the
trivial minimum
$|\rho_{\rm cdw}| = 0$.
This occurs at
$a = 3 \Tilde b^2/(16 d)$
if
$\Tilde b < 0$.

At arbitrary sign of
$\Tilde b$
the transition temperature can be compactly expressed as
\begin{eqnarray}
T_{\rm CDW} = T_* + \frac{ 3 (b-|c_4|)^2}{16 \alpha d} \theta(|c_4| - b).
\end{eqnarray} 
This shows that, unlike the
$n=3$
case, 
arbitrary weak 
$n=4$
umklapp term cannot change the continuous type of the transition. Only when
$|c_4|$
exceeds $b$, the transition becomes discontinuous. The point
$a=0$,
$|c_4| = b$
is a tricritical point on the phase diagram.
 
A version of the
$Z_4$
theory was used by the authors of
Refs.~\onlinecite{inagaki2018Z4, inagaki2019Erratum_toZ4},
who relied on a more complex Landau energy, with gradient terms, to
investigate possible CDW lock-in transition. It appears, however, that
their model was restricted by
$\tilde b > 0$
condition. Under such a constraint a sixth-order term in the Landau free
energy is not needed as the free energy remains stable for
$\Tilde b > 0$
even at fourth order. At the same time, one must remember that
the tricritical point cannot be reached unless
$\tilde b$
is allowed to pass through zero.

\subsection{$Z_5$ and $Z_6$ models}

For
$n=5$
and
$n=6$,
the phase diagram acquires additional complexity. We start our analysis by
writing the 
$n=5$
Landau free energy as
\begin{eqnarray}
\Tilde F^{(5)}
\! = \!
\frac{a}{2} |\rho_{\rm cdw}|^2 \!+\! \frac{b}{4} |\rho_{\rm cdw}|^4
\!-\!
\frac{|c_5|}{5} |\rho_{\rm cdw}|^5 \!+\! \frac{d}{6} |\rho_{\rm cdw}|^6,
\quad
\end{eqnarray} 
where, as before, we included the
$O(|\rho_{\rm cdw}|^6)$
term to provide proper growth of 
$\tilde F^{(5)}$
at 
$|\rho_{\rm cdw}| \rightarrow +\infty$.
Due to relative complexity of the
$n=5$
and
$n=6$
cases, it is convenient to introduce normalized quantities. Namely, the
dimensionless form of 
$\Tilde F^{(5)}$
reads
\begin{eqnarray}
\label{eq::n5_dimensionless_energy}
\frac{\Tilde F^{(5)}}{ {\cal F}_0} 
=
\frac{A}{2} y^2 + \frac{1}{4} y^4
-
\frac{C}{5} y^5 + \frac{1}{6} y^6.
\end{eqnarray} 
The coefficient $A$ in this formula is
\begin{eqnarray}
A = \frac{a d}{b^2} = \frac{\alpha d }{b^2}(T - T_*).
\end{eqnarray} 
Other quantities are
\begin{eqnarray} 
{\cal F}_0 = \frac{b^3}{d^2},
\quad
y = \sqrt{\frac{d}{b}}|\rho_{\rm cdw}|,
\quad
C = \frac{|c_5|}{\sqrt{bd}}.
\end{eqnarray} 
Here energy
${\cal F}_0$
sets the overall scale for
$\Tilde F^{(5)}$,
and $y$ is the dimensionless order parameter.
\begin{figure}[t!]
\centering
\includegraphics[width=0.99\columnwidth]{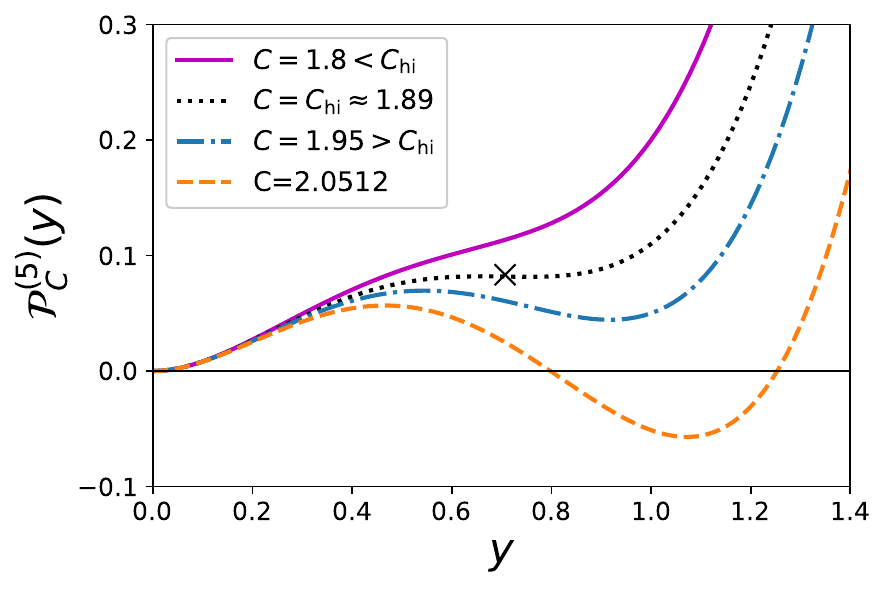}
\caption{Family of polynomials
${\cal P}^{(5)}_C (y)$,
for various $C$, see legend. When
$C < C_{\rm hi} = 4\sqrt{2}/3$,
the function increases monotonically for growing $y$ (solid magenta curve),
consequently,
Eq.~(\ref{eq::n5_phase_equation})
has one non-zero root if
$A<0$,
and no root otherwise. Exactly at
$C = C_{\rm hi}$
the polynomial graph (black dotted line) possesses a horizontal inflection
point at
$y=1/\sqrt{2}$,
which is marked by a cross. Below
$C_{\rm hi}$
the function is no longer monotonic (dashed and dash-dotted curves). In
this regime, for suitable $A$, multiple (two or three) roots of
Eq.~(\ref{eq::n5_phase_equation})
exist. For sufficiently large values of $C$ part of the curve lies below
horizontal axis, as the (blue) dashed curve demonstrates. In this case,
Eq.~(\ref{eq::n5_phase_equation})
has non-trivial roots even for positive $A$.
\label{fig::n5_polynomP}
}
\end{figure}
\begin{figure*}[t!]
\centering
\includegraphics[width=0.99\columnwidth]{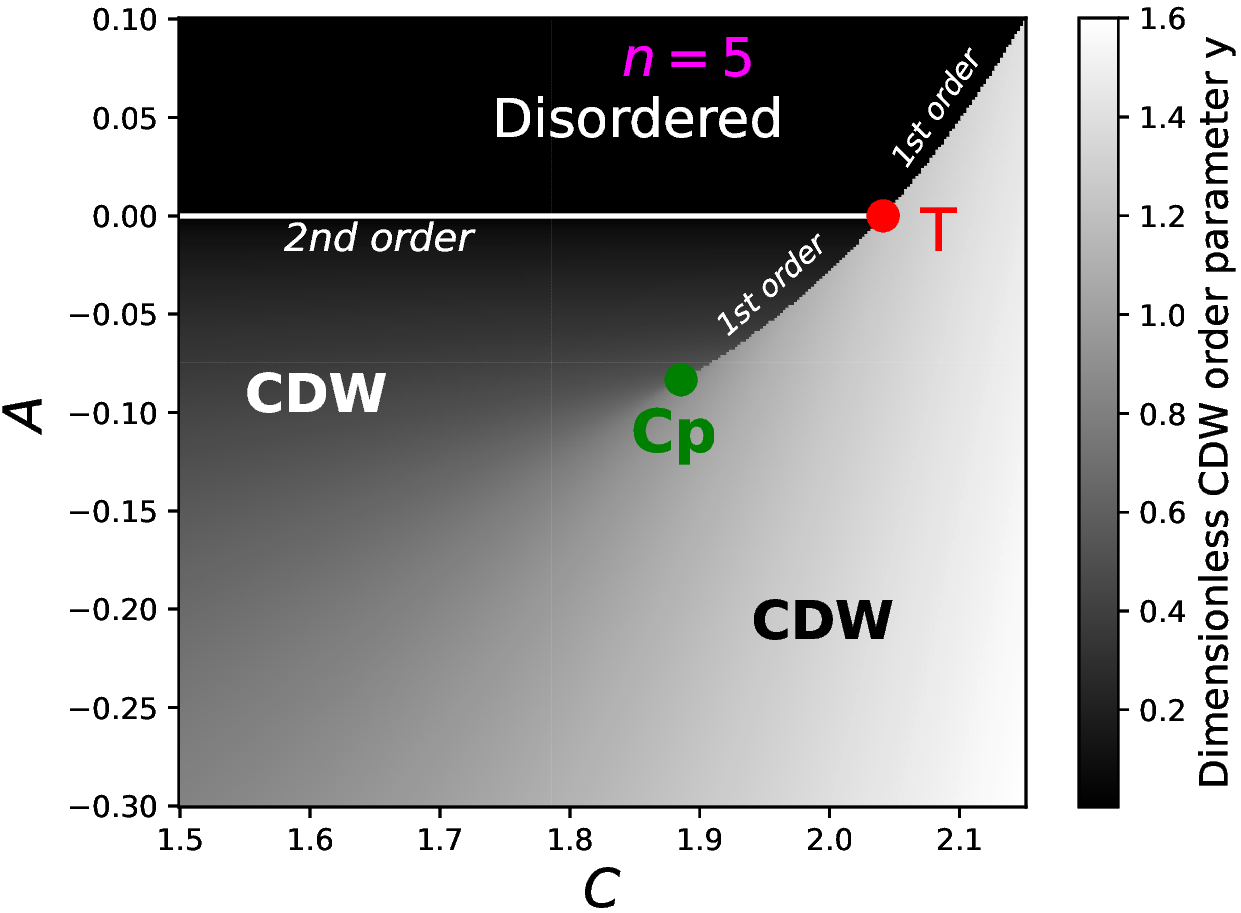}
\quad
\includegraphics[width=0.99\columnwidth]{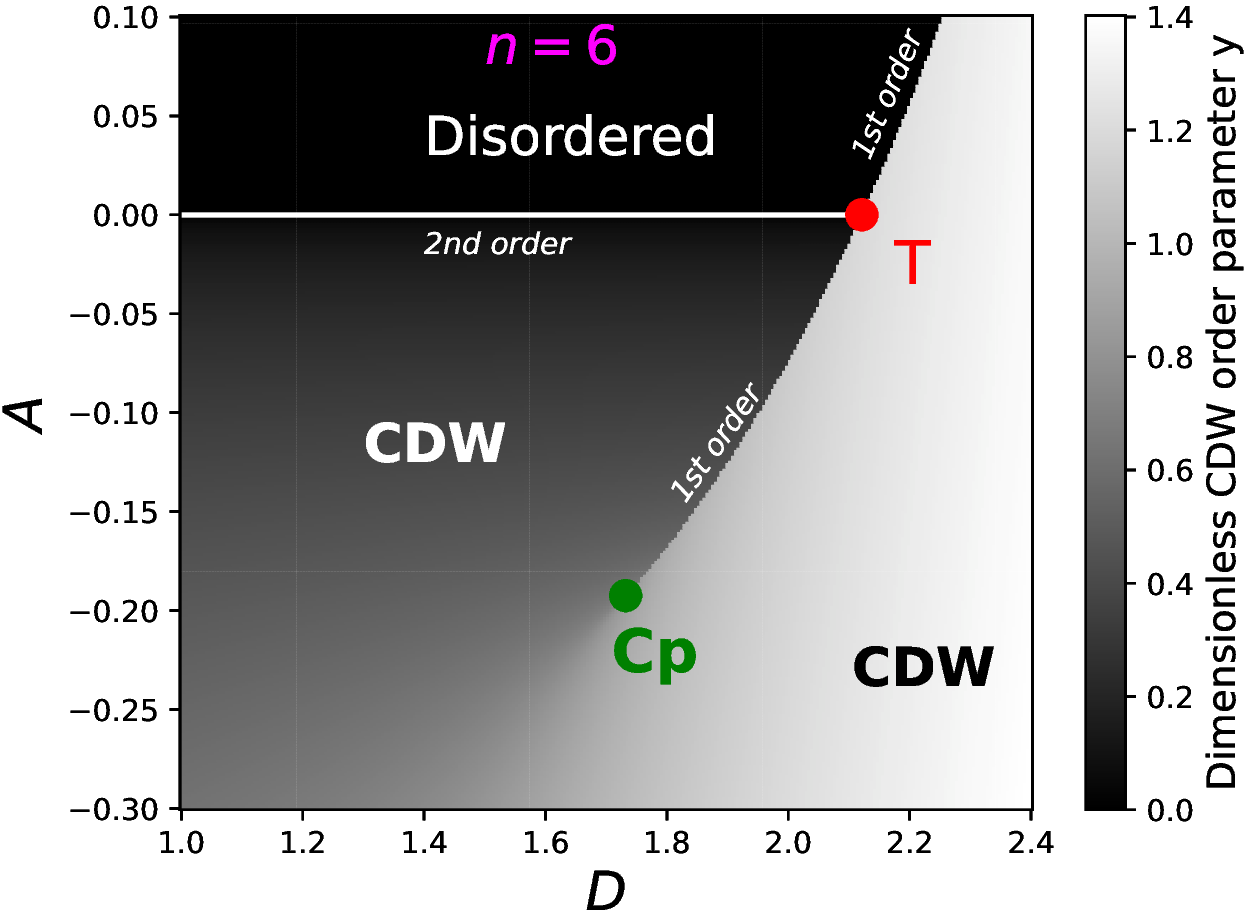}
\caption{Grayscale phase diagrams for
$n=5$
(left) and
$n=6$
(right). The diagrams are results of numerical minimization of the
dimensionless Landau free energies
$\Tilde F^{(5,6)}/{\cal F}_0$,
see
Eqs.~(\ref{eq::n5_dimensionless_energy})
and~(\ref{eq::n6_dimensionless_energy}),
over dimensionless order parameter $y$. Black area represents the
disordered phase, various shades of gray express the CDW order parameter
strength (for references, see a colorbar to the right of a respective phase
diagram). The disordered phase is bound on the right by a first-order
transition line visible as a sharp contrast edge. This first-order line
intrudes into the CDW phase terminating in a critical point (green dot
marked by `Cp'). Inside the ordered phase this line separates the CDW
states with unequal value of $y$ (crossing this line from left to right we
see discontinuous growth of $y$). The critical point `Cp' corresponds to
polynomial
${\cal P}^{(5,6)}_{C,D}$
with the horizontal inflection point [for
$n=5$
($n=6$)
this inflection occurs at
$C = 4\sqrt{2}/3 \approx 1.89$
(at
$D=\sqrt{3} \approx 1.73$)].
The second-order transition line at
$A=0$
limits the disordered phase from below. For
$n=5$
this line reaches the first-order transition curve at
$C=5/\sqrt{6} \approx 2.04$
and terminates there (this tricritical point is marked by a red dot and `T').
When
$n=6$,
point `T' is located at
$D=3/\sqrt{2} \approx 2.12$.
Qualitative structures of the two phase diagrams are identical.
\label{fig::n5_Pdiagram}
}
\end{figure*}
\begin{figure}
\centering
\includegraphics[width=0.99\columnwidth]{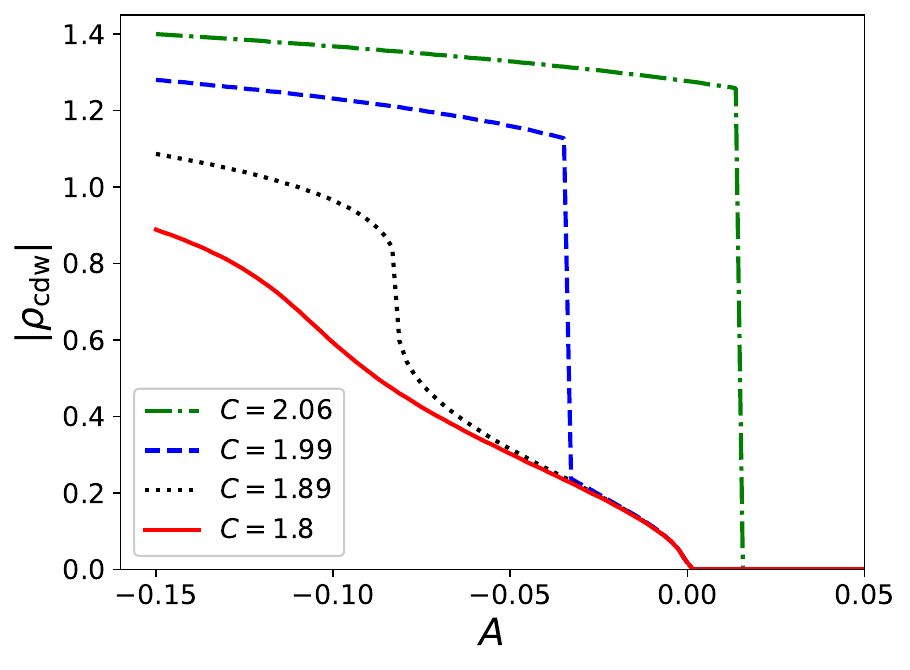}
\caption{Transition types in the
$Z_5$
model, for various values of $C$, see legend. For larger $C$, the
transition is first-order, as the (green) dash-dotted curve shows: when
temperature grows ($A$ increases) the order parameter 
$|\rho_{\rm cdw}|$
smoothly decreases until the transition point is reached, where
$|\rho_{\rm cdw}|$
drops to zero discontinuously. When
$C < 4\sqrt{2}/3$,
the transition is second-order, as the solid (red) curve plotted for
$C=1.8$
demonstrates. When
$4\sqrt{2}/3 < C < 5/\sqrt{6}$,
the model exhibits a cascade of two transitions, see (blue) dashed curve.
There is a continuous order-disorder transition at
$A=0$,
and a discontinuous transition within the same CDW phase at a lower
temperature. Finally, the (black) dotted curve corresponds to
$C=4\sqrt{2}/3$
(requires fine-tuning). As the system passes through `Cp' point in
Fig.~\ref{fig::n5_Pdiagram}\,(left),
a singularity
$\sim \pm |1/12 + A|^{1/3}$
emerges. The influence of this singularity on the solid (red) curve is
also visible.
\label{fig::rho_A}
}
\end{figure}

For 
$A<0$,
the disordered state
$y=0$
is absolutely unstable. It is at least metastable when
$A>0$.
As for ordered states (stable, metastable, or unstable), they are
represented by roots of the equation
\begin{eqnarray}
\label{eq::n5_phase_equation}
{\cal P}^{(5)}_C (y) = - A,
\end{eqnarray} 
where
${\cal P}^{(5)}_C (y)$
is a family of polynomials of variable $y$ 
\begin{eqnarray} 
{\cal P}^{(5)}_C (y) = y^2(1 - Cy + y^2),
\end{eqnarray} 
parameterized by
$C>0$.

For small $C$ and positive $y$, the polynomials are positive increasing
functions, see
Fig.~\ref{fig::n5_polynomP}.
Thus,
Eq.~(\ref{eq::n5_phase_equation})
has one solution for negative $A$. No solution exists when
$A>0$.
If $C$ is fixed, this describes an order-disorder continuous phase
transition at
$A=0$,
or, equivalently, at
$T=T_*$.

This simple picture is not applicable for
$C>C_{\rm hi} = 4\sqrt{2}/3$.
Indeed, at
$C=C_{\rm hi}$
a horizontal inflection at
$y = 1/\sqrt{2}$
is formed (hence, the subscript `hi'). For
$C > C_{\rm hi}$
the polynomial
${\cal P}^{(5)}_C (y)$
is no longer monotonic as a function of $y$, and more than one solution
become possible for appropriate (negative) values of $A$. Since
${\cal P}_ {C_{\rm hi}} (1/\sqrt{2}) = 1/12$,
multiple roots are realized for
$A>-1/12$.
Multiple non-trivial roots of
Eq.~(\ref{eq::n5_phase_equation})
implies that a first-order transition between CDW states emerges. Note that
the states separated by this transition have identical symmetries. The only
difference between these states is the magnitude of
$|\rho_{\rm cdw}|$,
which abruptly changes upon crossing the transition line.

As one can see from
Fig.~\ref{fig::n5_polynomP},
for sufficiently large $C$ there are finite intervals of $y$ in which the
value
${\cal P}^{(5)}_C (y)$
is negative. For such $C$,
Eq.~(\ref{eq::n5_phase_equation})
has two roots even for positive $A$. When $A$ grows, the roots approach
each other, merge, and ultimately disappear, signaling disappearance of a
(meta)stable CDW minimum.

The resultant phase diagram is shown in
Fig.~\ref{fig::n5_Pdiagram}~(left).
It features a second-order transition line reaching the first-order
transition curve. The latter terminates at a critical point inside the CDW
phase. This point corresponds to the horizontal inflection point for
${\cal P}^{(5)}_C$.
The location of the tricritical point `T', where two transition lines meet,
is characterized by the presence, at $A=0$, of a non-zero root of
Eq.~(\ref{eq::n5_phase_equation})
that additionally satisfies
$\Tilde F^{(5)} (y) = 0$.
These requirements are fulfilled when
$C=5/\sqrt{6}$,
which is the horizontal coordinate of `T'.

Depending on the value of $C$, the behavior
$\rho_{\rm cdw} = \rho_{\rm cdw} (T)$
may vary significantly, see
Fig.~\ref{fig::rho_A}.
If
$C > 5/\sqrt{6}$,
which corresponds to the area to the right of the `T' on the phase diagram,
the order-disorder transition is discontinuous. To the left of point `Cp'
($C<4\sqrt{2}/3$),
the transition is continuous, at
$A=0$.
In the interval
$4\sqrt{2}/3 < C < 5/\sqrt{6}$
the model exhibits a cascade of two transitions (a first-order CDW-CDW
transition followed by a second-order CDW-disorder transition). When
coefficient $C$ is fine-tuned to be
$C=4\sqrt{2}/3$,
the order parameter discontinuity shrinks to zero and becomes a continuous
singularity, as shown in
Fig.~\ref{fig::rho_A}.

For
$n=6$
commensuration, the Landau free energy can be expressed as
\begin{eqnarray}
\Tilde F^{(6)}
\! = \!
\frac{a}{2} |\rho_{\rm cdw}|^2 \!+\! \frac{b}{4} |\rho_{\rm cdw}|^4
\!-\! \frac{\tilde d}{6} |\rho_{\rm cdw}|^6
\!+\! \frac{e}{8} |\rho_{\rm cdw}|^8,
\quad
\end{eqnarray}
where
$\tilde d = |c_6| - d$,
and
$d, e > 0$.
Normalized form of this free energy is easy to establish
\begin{eqnarray}
\label{eq::n6_dimensionless_energy}
\frac{\Tilde F^{(6)}}{ {\cal F}_0} 
=
\frac{A}{2} y^2 + \frac{1}{4} y^4
-
\frac{D}{6} y^6 + \frac{1}{8} y^8.
\end{eqnarray} 
Here, under assumption
$b>0$,
we introduced the following set of parameters
\begin{eqnarray} 
y = \left(\frac{e}{b} \right)^{\!1/4} \!|\rho_{\rm cdw}|,
\quad
{\cal F}_0 = \frac{b^2}{e},
\\
A = a \sqrt{\frac{e}{b^3}},
\quad
D = \frac{\tilde d}{\sqrt{b e}}.
\end{eqnarray} 
Similar to 
Eq.~(\ref{eq::n5_phase_equation}),
ordered phases of the
$n=6$
model are represented by roots of equation
\begin{eqnarray}
{\cal P}^{(6)}_D (y) = -A,
\quad
\text{where}
\quad
{\cal P}^{(6)}_D = y^2 ( 1 - D y^2 + y^4).
\end{eqnarray} 
Analysis of
Eq.~(\ref{eq::n5_phase_equation})
can be adopted for the latter equation, and an
$n=6$
phase diagram can be constructed, see
Fig.~\ref{fig::n5_Pdiagram}~(right).
It is clear that both diagrams in
Fig.~\ref{fig::n5_Pdiagram}
are qualitatively similar. 

The
$Z_5$
model can be used to describe the first-order transition in
IrTe$_2$.
Specifically, at
$T \sim 280$\,K
the compound
enters~\cite{IrTe2_prb2014}
a CDW phase with
${\bf k} = (1/5, 0, 1/5)$.
Obviously, this phase fits the
$Z_5$
case. Observed hysteresis and resistivity discontinuities indicate a
first-order transition. Since no incommensurate order was reported, the
lock-in scenario can be excluded. This implies relevance of the phase
diagram in
Fig.~\ref{fig::n5_Pdiagram}\,(left).

At lower temperature the
$n=5$
phase in 
IrTe$_2$
is replaced by a
$n=8$
CDW state. The transitions are sensitive to pressure and Ir-to-Pt chemical
substitution. Thus, one can consider constructing a more comprehensive
phase diagram that accounts for these circumstances. Such a deserving task,
however, must be deferred to future research.

As for the
$Z_6$
model, it will be of use for the discussion of
EuTe$_4$,
see next section.

\section{Near-commensurate CDW}
\label{sec::near_commensurate}

We demonstrated in the previous section that symmetry-allowed umklapp terms
enhance complexity of the model's phase diagram. In particular, a
first-order transition line emerges. Let us now generalize our approach to
the case of NC-CDW. 

For an NC-CDW, vector
$n {\bf k}$
does not belong to the reciprocal lattice of a host crystal for any
$n \in \mathbb{N}$,
however, one can find a (small) integer $m$ and a reciprocal lattice vector 
${\bf b} \ne 0$
such that a ``defect'' vector
\begin{eqnarray} 
\label{eq::near_commensurate_def}
{\bf p} = {\bf b} - m {\bf k}
\end{eqnarray} 
is small in the sense that
$|( {\bf p} \cdot {\bf a}_i )| \ll 1$
for all
$i=1,2,3$.

Since
${\bf k}$
is not commensurate,
$\rho_{\rm cdw}^n$
is not compatible with the lattice translation group for any
$n \in \mathbb{N}$.
Yet, an umklapp contribution associated with the NC-CDW order can emerge
through the following mechanism. Note that a monomial
$\rho_{\rm cdw}^m$,
where $m$ is defined in
Eq.~(\ref{eq::near_commensurate_def}),
transforms according to the rule
$\rho_{\rm cdw}^m \rightarrow e^{- i ({\bf p} \cdot {\bf a}_i)}
\rho_{\rm cdw}^m$
upon a translation on the elementary lattice vector
${\bf a}_i$.
Although ``the elementary defects"
$e^{- i ({\bf p} \cdot {\bf a}_i)}$
are close to unity, the exponent oscillates for longer translations,
indicating that the contribution
$\propto \rho_{\rm cdw}^m$
averages to zero upon summation over the whole sample. 

Fortunately, since
$|{\bf p}|$
is small, the lattice can adjust its structure to allow the umklapp term.
Imagine that the lattice, in response to the CDW presence, experiences an
additional periodic distortion with the wave vector 
${\bf p}$.
Representing such a distortion by a complex quantity $u$, one can prove
that the term
$u \rho_{\rm cdw}^m$
is invariant under the lattice translations. Indeed, a translation on 
${\bf a}_i$
transforms $u$ according to the rule
$u \rightarrow e^{i ({\bf p} \cdot {\bf a}_i)} u$.
This makes the products
$u \rho_{\rm cdw}^m$
and
$(u \rho_{\rm cdw}^m)^*$
translation-invariant, and admissible contributions to the Landau free
energy.

Conceptualizing $u$, one can think of it as a ``frozen", or ``condensed"
phonon mode, whose wave vector is
${\bf p}$.
More detailed analysis of the origin of $u$ can be found in
Appendix~\ref{app::NCCD}.
There we formulate a specific model that assigns a very concrete meaning to
$u$. In addition, we discuss a broader context within which the notion of
$u$ may be justified.

With this in mind, we write the following Landau-type model
\begin{eqnarray}
\label{eq::F_NC0}
F_{\rm NC} (\rho_{\rm cdw}, u)
&=&
F_0 (\rho_{\rm cdw})
\\
\nonumber 
&+&
\kappa_{{\bf p}} |u|^2
+
\frac{1}{\sqrt{2m}}
\left( g_{\rm nc} u \rho_{\rm cdw}^m + {\rm c.c.} \right).
\end{eqnarray} 
Here
$\kappa_{{\bf p}} |u|^2 \geq 0$
is the elastic energy associated with the distortion $u$, complex
coefficient 
$g_{\rm nc}$
is a coupling constant, and factor
$(2m)^{-1/2}$
is introduced into this formula to make expression below consistent with
previous definitions. (A more detailed discussion of the motivation behind 
Eq.~(\ref{eq::F_NC0})
can be found in
Appendix~\ref{app::NCCD}.)

Minimizing this energy over
$u^*$,
we obtain
\begin{eqnarray}
\label{eq::u_minimum}
u
=
- \frac{g_{\rm nc}^*}{\sqrt{2m} \kappa_{{\bf p}}}
 (\rho_{\rm cdw}^*)^m.
\end{eqnarray}
Substituting the relation for $u$ into
Eq.~(\ref{eq::F_NC0})
one derives the reduced free energy that depends on
$|\rho_{\rm cdw}|$
only
\begin{eqnarray}
\label{eq::F_NC}
\Tilde F_{\rm NC}^{(m)} 
=
F_0 
-
\frac{|c_{2m}|}{2m} |\rho_{\rm cdw}|^{2m},
\ 
\text{where}
\ 
|c_{2m}| = \frac{|g_{\rm nc}|^2}{\kappa_{{\bf p}}}.
\end{eqnarray} 
Due to vector
${\bf p}$
being small in an NC-CDW phase, the corresponding stiffness 
$\kappa_{\bf p}$
is small, as
Appendix~\ref{app::stiffness}
demonstrates. Thus, coefficient 
$|c_{2m}|$
may be significant.

Note that this reduced Landau free energy is identical to
$\Tilde F^{(n)} (|\rho_{\rm cdw}|)$
for 
$n=2m$.
That is,
$\Tilde F_{\rm NC}^{(m)} \equiv \Tilde F^{(2m)}$.
Therefore,
$\Tilde F_{\rm NC}^{(m)}$
can describe a first-order transition between the disordered and ordered
phases, similar to what we have demonstrated above for
$Z_4$
and
$Z_6$
models. Beside this, for
$m=3$,
a first-order transition within the same CDW phase, as visible in the
$Z_6$
model phase diagram (see
Fig.~\ref{fig::n5_Pdiagram}),
can be realized.

In connection with the latter possibility, we would like to discuss briefly
the case of
EuTe$_4$,
see
Refs.~\onlinecite{EuTe4hyster2019exper,EuTe4hysteresis2022exper,
EuTe4hysteresis2023exper}.
There are at least two NC-CDW order parameters, one of them can be
described~\cite{EuTe4hyster2019exper}
as being nearly-commensurate
${\bf k} \approx {\bf b}/m$,
with
$m=3$.
Inside the ordered phase a temperature-driven hysteresis loop that extends
from 80\,K to at least 400\,K was observed. A lock-in transition behind
this hysteresis was ruled
out~\cite{EuTe4hysteresis2022exper,lv2025EuTe4},
since all wave vectors associated with the CDW order demonstrate remarkable
thermal stability, and no commensurate diffraction peaks.

On the other hand, we can apply 
$\Tilde F_{\rm NC}^{(3)}$
to this system. Since
$\Tilde F_{\rm NC}^{(3)}$
is identical to
$\Tilde F^{(6)}$,
the two have the same phase diagram, see
Fig.~\ref{fig::n5_Pdiagram}\,(right).
There, let us focus on the first-order transition line between `Cp' and
`T': when this line is crossed, the amplitude
$|\rho_{\rm cdw}|$
discontinuously changes between two non-zero values. This first-order
transition inside the CDW order may be the source of the hysteretic
behavior of
EuTe$_4$.

\section{Discussion}
\label{sec::discussion}

We demonstrated above that the umklapp contributions to the CDW Landau free
energy can qualitatively affect the phase diagram of a CDW-hosting
material. Our approach is based on the standard McMillan-Nakanishi-Shiba
framework, and represents an extension of this framework appropriate for a
unidirectional commensurate or nearly-commensurate CDW.

We saw that the effects introduced by the umklapp term depend on the
commensuration degree $n$: the larger $n$ the richer the model's phase
diagram. Indeed, for 
$n=2$
the umklapp contribution does nothing but shifts the transition point, when
$n$ is as large as 5 or 6, the phase diagram displays such elements as
tricritical point, critical point, continuous and discontinuous transitions
lines, see
Fig.~\ref{fig::n5_Pdiagram}.

Moreover, the
$Z_{5,6}$
models allow for a possibility that the destruction of the order may occur
through a two-step process: lower-temperature CDW-CDW discontinuous
transition followed by higher-temperature CDW-disorder continuous
transition, as 
Fig.~\ref{fig::rho_A}
illustrates for
$Z_5$
model. Superficially, one may argue that such a cascade was already
discussed quite some time ago (see, for instance,
Fig.~1 
in
Ref.~\onlinecite{TaSe2McMillan1977microscopic_theory},
or
Fig.~2 
in
Ref.~\onlinecite{nakanishi1977domain}).
However, there is an important difference. Indeed, in our case, the
first-order CDW-CDW transition occurs within the same commensurate phase.
This is very much unlike commensurate-incommensurate CDW lock-in
transitions of
Refs.~\onlinecite{TaSe2McMillan1975theory_umklapp,
TaSe2McMillan1977microscopic_theory},
as well as
other~\cite{nakanishi1977domain}
first-order transitions associated with discontinuous change of CDW wave
vector. 

We did not extend our analysis beyond
$n=6$
commensuration degree. Unfortunately, we were unable to identify any
general principle restricting structure of the phase diagram of large-$n$
models. In such a situation any investigation of a large-$n$ case becomes
problematic due to ever increasing number of parameters one must keep in
the Landau free energy expansion to guarantee its stability. Yet, apart
from these purely technical issues, the formulated analytical framework can
be adopted to
$n>6$
models.

Our argumentation can be extended to NC-CDW order parameters as well. This
is not ultimately that surprising: in a situation of small deviation
from commensurability
$|{\bf p}|$
at not-too-large $n$ a sufficiently soft hosting crystal lattice
reorganizes itself to lock-in with the CDW. This is the heuristic
understanding behind
Eq.~(\ref{eq::F_NC}).

This formalism can be used to explain the observed hysteretic
behavior~\cite{EuTe4hyster2019exper,EuTe4hysteresis2022exper,
EuTe4hysteresis2023exper}
of the CDW order in
EuTe$_4$.
The stability of the measured CDW wave vector and no commensurate
diffraction peaks at any temperature are inconsistent with the lock-in
CCDW/NC-CDW mechanism. On the other hand, a NC-CDW/NC-CDW transition with
unchanging
${\bf k}$
can be explained by the model in
Sec.~\ref{sec::near_commensurate}. 
Since the CDW in
EuTe$_4$
is close to
$m=3$
commensuration,
Eq.~(\ref{eq::F_NC})
indicates that
$Z_6$
model may be relevant for this compound.

More ``microscopic" discussion motivating the applicability of
Eq.~(\ref{eq::F_NC0})
to
EuTe$_4$
and, possibly, other layered materials hosting NC-CDW can be found in
Appendix~\ref{app::NCCD}.
At the same time, one must remember that the CDW phase in
EuTe$_4$
is quite complicated. The compound demonstrates two co-existing
unidirectional order parameters with non-identical wave vectors. They
interact with each other and with the lattice. Clearly, proper theoretical
understanding of this complexity requires dedicated research efforts, and
beyond the scope of this paper. It is also worth mentioning that
Ref.~\onlinecite{EuTe4hysteresis2022exper}
offers alternative explanation to the origin of the first-order transition in
EuTe$_4$.

We always assumed above that the system invariably chooses the global
minimum of its Landau free energy. Yet for
$n>2$
our models allows for metastable states. For example, metastable minima,
representing both ordered and disordered states, are clearly visible in
Fig.~\ref{fig::n3},
which is plotted for
$n=3$.
Delayed departure from a metastable minimum reveals itself as hysteresis, a
common fixture of experimental presentation of a first-order transition.
Using the above phase diagrams for experimental data analysis one must
remember that our calculations do not take hysteresis into account. In
principle, hysteresis can be captured in the framework of the Landau theory
of phase transitions. However, the description of this kind oversimplifies
the physics significantly as it ignores various non-universal mechanisms
affecting hysteretic behavior in real materials.

The discussed ideas can be adopted to the spin-density wave (SDW) case as
well. The SDW order parameter is a complex vector
${\bf S}$,
and
$({\bf S}\cdot {\bf S})$
is a true complex scalar. Thus, for even integer 
$n=2k$
one can construct an umklapp term of the form
$c_{2k} ({\bf S}\cdot {\bf S})^k + {\rm c.c.}$
that is consistent with discrete translations, parity, and time inversion
symmetries.

Finally, let us make the following observation. A number of alloys
demonstrate the first-order transition between disordered and CDW phases.
Several papers reporting this also
commented~\cite{Lu5Ir4Si10first_orderCDW1999exper,
Lu2Ir3Si5_cdw2006hysteresis_exper,
Lu2Ir3Si5_first_orderCDW2015exper}
that such an unusual transition type must be a consequence of ``strong
coupling". Within the context of our formalism the expectation of strong
coupling regime is quite natural: the large-$n$ umklapp coefficients 
$|c_n|$
are likely to be small unless the lattice modulations associated with the
order parameter are
significant~\cite{LEE1974703,pnas2012umklapp}.

This suggests two things. Firstly, in order to observe the ``non-trivial"
features of the phase diagrams, such as the critical and tricritical
points, and the first-order transition line, we should search among
commensurate or nearly-commensurate CDW-hosting materials that demonstrate
pronounced amplitude of the CDW modulations. A possible tool to control
these amplitudes is through chemical substitution. The latter indeed can
exert powerful influence on a CDW phase, as illustrated by the
$R$Te$_3$
series whose CDW features vary dramatically as the rare-earth atom $R$
changes~\cite{chem_pressure2014tritellurides}.

Secondly, since in the systems of interest the order parameter amplitude
must be significant, weak-coupling approximations, with their analytical or
semi-analytical prescriptions for the Landau free energy coefficients,
become of questionable accuracy. Thus, one must exclusively employ
numerical material-science methods to extract the Landau expansion. In this
regard we can cite
Ref.~\onlinecite{dft_landau2014artyukhin}
which found the coefficients for a multiferroic material. Adopting this
program for CCDW-hosting materials appears to be a useful direction for
future research.

To conclude, in this paper, within the Landau free energy framework, we
explored effects of the order parameter commensuration on the phase diagram
of a CDW-hosting system. We demonstrated that in the case of commensurate and
nearly-commensurate CDW the anticipated second-order transition may be
replaced by the first-order transition, as indeed observed experimentally.
Under certain circumstances our model predicts a cascade of two transitions
(low-temperature CDW-CDW first-order transition is followed by
higher-temperature order-disorder second-order transition). These ideas may
be applicable to SDW phases as well.

\section*{Acknowledgments}

Author is thankful to B.V.~Fine, B.Q.~Lv, and Alfred Zong for illuminating
discussions.

\appendix

\section{Model of a nearly-commensurate CDW in a layered crystal}
\label{app::NCCD}

\begin{figure}[t!]
\centering
\includegraphics[width=0.99\columnwidth]{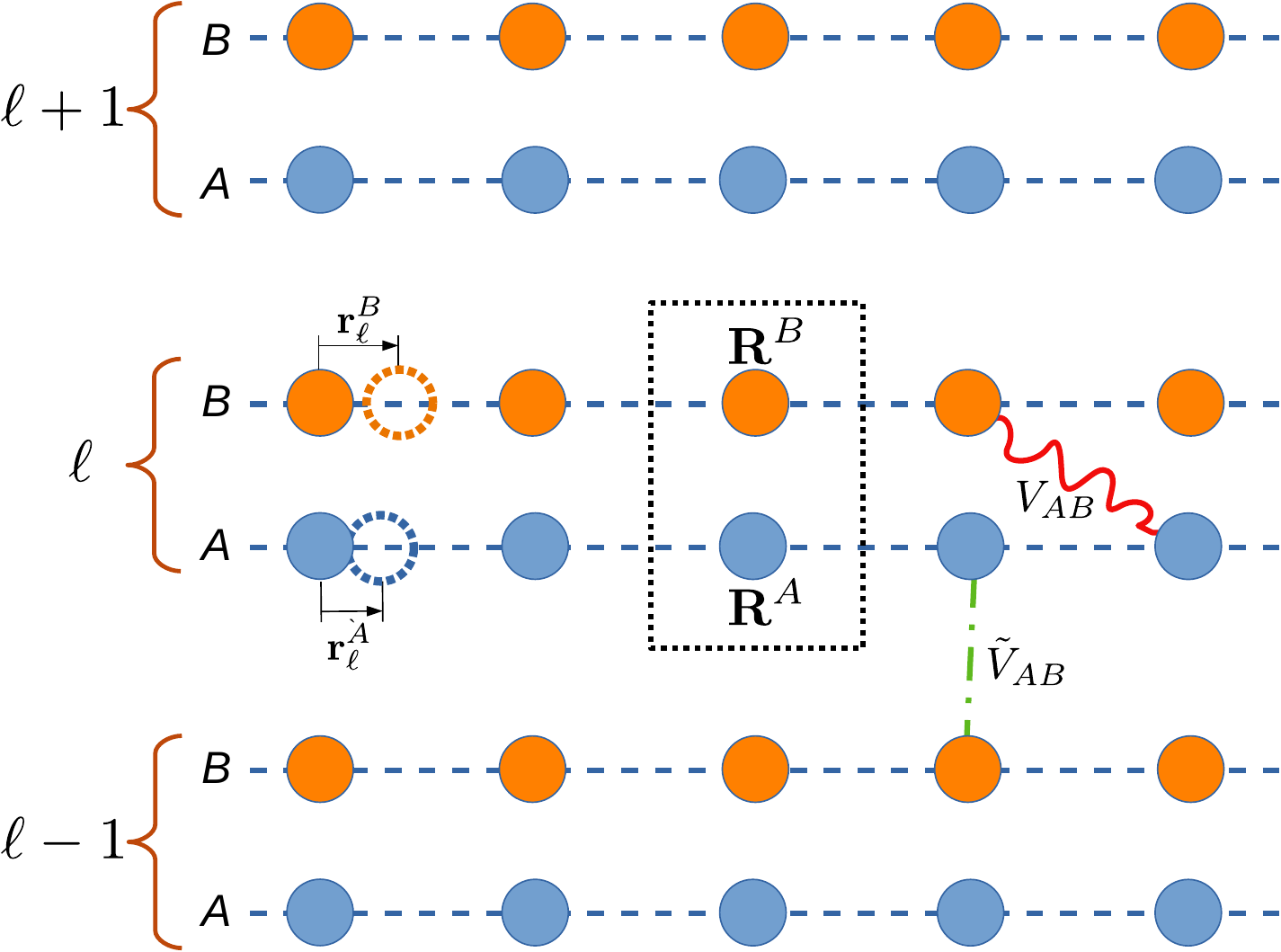}
\caption{
Sketch of the quasi-2D crystal lattice discussed in 
Appendix~\ref{app::NCCD}.
The lattice is a stack of 
$\cal N_\perp$
weakly coupled 2D units. Each unit itself is composed of two atomic planes,
such that plane~B is placed on plane~A. Lattice coordinates inside the
planes are denoted as
${\bf R}^{A,B}$.
(For simplicity, the structure of 2D lattices in A and B is assumed to be
the same.) Integer index
$\ell = 1, \cdots, {\cal N}_\perp$
labels the 2D units and serves as a lattice coordinate orthogonal to the
planes. An elementary cell is shown as a dotted-line rectangle. It contains
one atom from plane~A (blue circle) and one atom from plane~B (orange
circle). The CDW instability is localized in planes~A of each unit, while
planes~B experience distortions in response to the CDW. Intra-unit
interaction 
$V_{AB}$
is represented by (red) wavy curve, while inter-unit interaction 
$\Tilde V_{AB}$
corresponds to (green) dash-dotted line. Atoms positions distorted by the
CDW are described by vectors
${\bf r}^{A,B}_\ell$.
\label{fig::model_lattice}
}
\end{figure}

Here we describe a model in which coupling between NC-CDW and a non-CDW
lattice degree of freedom give rise to the Landau
energy~(\ref{eq::F_NC0}).
We consider a stack of
${\cal N_\perp}$
weakly coupled two-dimensional (2D) units forming a quasi-2D
three-dimensional body, see
Fig.~\ref{fig::model_lattice}.
Each unit itself is composed of two 2D lattices of non-identical chemical
composition (every lattice contains
${\cal N}_{\rm 2D}$
sites). One of these two lattices, denoted below as plane~A, hosts a CDW
instability. The other lattice (plane~B) is stable by itself, but it is
coupled to plane~A atoms by a short-range interaction. Thus, CDW
distortions, which originate in plane~A, cause deformations in plane~B as
well.

While we do not attempt to capture the behavior of a specific material, it
is worth noting that various tellurides of rare-earth elements share
important similarities with such a model. For example,
EuTe$_4$
can be viewed as a stack of flat Te~planes separated by corrugated EuTe
layers. The CDW is hosted by the Te~layers, while EuTe planes are believed
to be passive
spacers~\cite{EuTe4hysteresis2022exper}.

We start formal description of our model by writing down the free energy as
a sum of three terms
\begin{eqnarray}
F_{\rm q2D} = F_A + F_B + F_{AB},
\end{eqnarray} 
where
$F_A = F_0$
describes the CDW instability inside planes~A. The free energy 
$F_B$
for planes~B describes a stable lattice, see
Appendix~\ref{app::stiffness}.
Finally, the term
$F_{AB}$
represents interaction between A and B. It reads
\begin{eqnarray}
F_{AB}
=
\sum_{ {\bf R}^A {\bf R}^B \ell}
	V_{AB} \left( 
		{\bf R}^A + {\bf r}_\ell^A - {\bf R}^B - {\bf r}_\ell^B
	\right)
	+
\\
\nonumber 
	\tilde V_{AB} \left( 
		{\bf R}^A + {\bf r}_\ell^A - {\bf R}^B - {\bf r}_{\ell+1}^B
	\right).
\end{eqnarray} 
Integer-valued index $\ell$ counts 2D units, while within each unit
summation over every site 
${\bf R}^A$ 
in plane~A and every site
${\bf R}^B$
in plane~B is executed. (We assume that plane~A lattice is the same as
plane~B lattice.) Displacement of an atom from its position
${\bf R}^A$
in plane~A 
of unit $\ell$ is expressed by vector
${\bf r}^A_\ell= {\bf r}^A_\ell({\bf R}^A)$,
see
Fig.~\ref{fig::model_lattice}.
Similarly, displacement of an atom in plane~B is expressed by vector
${\bf r}^B_\ell= {\bf r}^B_\ell({\bf R}^B)$.
Short-range interactions energies 
$V_{AB}$
and
$\Tilde V_{AB}$
account for intra-unit and inter-unit interactions between the atoms
constituting the crystal.

Under assumption of the smallness of
$|{\bf r}^{A,B}_\ell|$
relative to 2D lattice constant, Taylor expansion 
$F_{AB} = F_{AB}^0 + \delta F$
is performed, where
\begin{eqnarray} 
F_{AB}^0
=
{\cal N}_\perp
{\cal N}_{\rm 2D}
\sum_{ {\bf R}^A}
	V_{AB} \left( {\bf R}^A \right)
	+
	\tilde V_{AB} \left( {\bf R}^A \right)
\end{eqnarray}
is the energy of the undistorted lattice, while the correction
$\delta F$
due to the distortions is 
\begin{eqnarray}
\label{eq::DE}
\delta F
=
\sum_{ {\bf R}^A {\bf R}^B \atop \ell n}
	\frac{1}{n!}
	\left[ ({\bf r}_\ell^A - {\bf r}_\ell^B) \cdot \nabla \right]^{n}
	V_{AB} \left( {\bf R}^A - {\bf R}^B \right)
\\
\nonumber 
	+
	\frac{1}{n!}
	\left[ ({\bf r}_\ell^A - {\bf r}_{\ell+1}^B) \cdot\nabla \right]^n
	\tilde V_{AB} \left( {\bf R}^A - {\bf R}^B \right).
\end{eqnarray}
Here the gradient operators
$\nabla$
differentiate 
$V = V( {\bf R} )$
and
$\Tilde V = \Tilde V( {\bf R} )$,
and do not affect
${\bf r}^{A,B}_\ell 
=
{\bf r}^{A,B}_\ell ({\bf R})$.

To this moment we did not assume anything specific about the distortions
${\bf r}_\ell^{A,B}$.
Now let us imagine that the planes~$A$ distortions are associated with the
CDW state
\begin{eqnarray}
{\bf r}_\ell^A
=
\frac{\rho_{\rm cdw} {\bf k}}{|{\bf k}|^2}
e^{i {\bf k} \cdot {\bf R}^A } + {\rm c.c.}
\end{eqnarray} 
The distortion here is chosen to be ``longitudinal'', such that
$\nabla \cdot {\bf r}_\ell^A
=
i \rho_{\rm cdw} e^{i {\bf k} \cdot {\bf R} } + {\rm c.c.}$
is the (dimensionless) CDW order parameter, representing particle number
per unit cell.

As planes~B react passively to the CDW, displacements
${\bf r}_\ell^B$
are expected to be weaker than those in planes~A:
$|{\bf r}_\ell^B| \ll |{\bf r}_\ell^A|$.
Due to this smallness, we will neglect those terms in
$\delta F$
that are second- and higher-order in
$|{\bf r}_\ell^B|$.
In this limit, we can assume simple trigonometric dependence
\begin{eqnarray}
\label{app::passive_u_def}
{\bf r}_\ell^B ({\bf R}^B)
=
\frac{u {\bf k}}{|{\bf k}|} e^{ i {\bf p} \cdot {\bf R}^B } + {\rm c.c.},
\end{eqnarray} 
where $u$ is a complex scalar. For now, the value of the wave vector 
${\bf p}$
remains unconstrained. It will be determined below.

We can approximately express 
$\delta F$
as a sum
$\delta F \approx \delta F^{(0)} + \delta F^{(1)}$,
where
$\delta F^{(0)}$
is independent of
${\bf r}_\ell^B$,
and
$\delta F^{(1)} = O( u )$.
Of these two terms,
$\delta F^{(0)}$
introduces renormalizations to 
$F_A$
due to interaction between the CDW order parameter in planes~A and
undistorted lattice in planes~B. We assume that all such renormalizations
are already included in
$F_A$,
and will ignore
$\delta F^{(0)}$.

As for
$\delta F^{(1)}$,
it is responsible for coupling between the CDW and plane~B lattice
deformation. To find
$\delta F^{(1)}$,
we perform two substitutions:
${\bf R}^{A} \rightarrow {\bf R}^A + {\bf R}^B$,
and
$n \rightarrow n+1$.
Then, introducing new notation
\begin{eqnarray}
\tilde {\bf r}_\ell^A 
=
\frac{\rho_{\rm cdw} {\bf k}}{|{\bf k}|^2}
e^{i {\bf k} \cdot ({\bf R}^A + {\bf R}^B) } + {\rm c.c.},
\end{eqnarray}
we can write
\begin{eqnarray}
\label{eq::DE_shifted}
\delta F^{(1)}
=
\!-\!\! \sum_{ {\bf R}^A {\bf R}^B \atop \ell n}
	\frac{1 }{n!}
	(\Tilde {\bf r}_\ell^A \cdot \nabla)^{n} 
	({\bf r}_\ell^B \cdot \nabla )
	V_{AB} \left( {\bf R}^A \right)
	+
\\
\nonumber 
	\frac{1 }{n!}
	(\Tilde {\bf r}_\ell^A \cdot \nabla)^{n} 
	({\bf r}_{\ell + 1}^B \cdot \nabla )
	\Tilde V_{AB} \left( {\bf R}^A \right),
\end{eqnarray} 
where summation over $n$ starts from
$n=0$,
and operator $\nabla$ acts on functions 
$V_{AB}$
and
$\Tilde V_{AB}$
only [see explanation below
Eq.~(\ref{eq::DE})].
 
Note that lattice variable
${\bf R}^B$
enters 
Eq.~(\ref{eq::DE_shifted})
only through complex exponents of the form
$e^{\pm i {\bf K} \cdot {\bf R}^B}$,
where
${\bf K} = m {\bf k} \pm {\bf p}$,
and $m$ is a non-negative integer. In such a situation, summation over
${\bf R}^B$
can be executed explicitly with the help of the ``quasi-momentum
conservation rule": if
${\bf K}$
belongs to the reciprocal lattice, then
\begin{eqnarray} 
\label{eq::qmomentum_conserv}
\sum_{{\bf R}^B} e^{ \pm i {\bf K}\cdot {\bf R}^B} = {\cal N}_{\rm 2D}.
\end{eqnarray} 
Otherwise, the sum is zero.

Applying this selection rule to 
Eq.~(\ref{eq::DE_shifted}),
one finds that summation over
${\bf R}^B$
produces non-zero results only when
${\bf p} = {\bf b} \pm m {\bf k} $,
where 
${\bf b}$
is a reciprocal lattice vector. The terms that survive the summation are
${\sim u \rho_{\rm cdw}^m}$,
as well as 
${\sim u \rho_{\rm cdw}^n}$,
$n>m$.
They introduce umklapp coupling between the CDW in planes~A and the lattice
degrees of freedom in planes~B. This derivation may be viewed as a
justification for
Eq.~(\ref{eq::F_NC0}).

Straightforward analysis of this procedure for a generic incommensurate
CDW, with small inter-plane couplings and sufficiently stiff planes~B,
demonstrates that the interactions between the CDW and the passive
structures of the lattice introduce quantitative corrections only. Indeed,
weak inter-plane coupling means that
$g_{\rm nc}$
in
definition~(\ref{eq::F_NC})
for
$|c_{2m}|$
is small, while stiff planes~B implies that
$\kappa_{{\bf p}}$
in the same expression is large. These two factors act together to suppress
$|c_{2m}|$,
making sure that the umklapp contribution does not introduce qualitative
modifications to the phase diagram.

This simple conclusion, however, must be reconsidered for an NC-CDW
phase. In this case, the coupling term satisfying
condition~(\ref{eq::near_commensurate_def})
is of particular importance. Since ``the defect" of the NC-CDW state is
small
$|{\bf p}| \ll \pi/a$,
therefore, the stiffness
$\kappa_{{\bf p}}$,
which controls the plane~B distortion with the wave vector
${\bf p}$,
is small as well (see
Appendix~\ref{app::stiffness}).
Consequently, the coefficient 
$|c_{2m}|$
in
Eq.~(\ref{eq::F_NC})
is large due to small denominator.

In connection with the latter observation, let us make a clarification. It
appears as if 
$|c_{2m}|$
can become arbitrary large for very small 
$|{\bf p}|$.
We must remember, however, that for very small
$\kappa_{{\bf p}}$
the value of $u$ in
Eq.~(\ref{eq::u_minimum})
becomes very large. As the whole formalism is based on $u$ being small,
the limit of very small
$\kappa_{{\bf p}}$
should be treated by subtler approaches that account for higher-order terms
$u^s$,
$s>1$.
Also, in more sophisticated treatment, inter-layer interactions must be
accounted for.

Finally, we can evaluate
$g_{\rm nc}$
introduced in
Eq.~(\ref{eq::F_NC0}).
The calculations below are performed for a specific case of
$m=3$,
which means that
$3{\bf k}$
is ``almost" a reciprocal lattice vector. Adaptation for higher $m$ are
cumbersome but straightforward.

We start by isolating the
$O(|\rho_{\rm cdw}|^3)$
term 
$\delta f^{(3)}$
in
Eq.~(\ref{eq::DE_shifted})
and write
\begin{widetext}
\begin{eqnarray}
\delta f^{(3)}
=
- {\cal N}_\perp \frac{u\rho_{\rm cdw}^3}{6 |{\bf k}|^7}
\sum_{ {\bf R}^A {\bf R}^B }
	e^{3 i {\bf k} \cdot ({\bf R}^A + {\bf R}^B) 
		+ i {\bf p} \cdot {\bf R}^B}
	({\bf k} \cdot \nabla)^{4} 
	\left[
		V_{AB} \left( {\bf R}^A \right)
		+
		\Tilde V_{AB} \left( {\bf R}^A \right)
	\right]
+
{\rm c.c.}
\end{eqnarray} 
Since
$3 {\bf k} + {\bf p}$
is a reciprocal lattice vector, summation over
${\bf R}^B$
may be executed using
Eq.~(\ref{eq::qmomentum_conserv}).
Consequently
\begin{eqnarray}
\delta f^{(3)}
=
- {\cal N}_\perp {\cal N}_{2D}
\frac{ u \rho_{\rm cdw}^3}{6 |{\bf k}|^7}
\sum_{ {\bf R}^A }
	e^{ - i {\bf p} \cdot {\bf R}^A }
	({\bf k} \cdot \nabla)^{4} 
	\left[
		V_{AB} \left( {\bf R}^A \right)
		+
		\Tilde V_{AB} \left( {\bf R}^A \right)
	\right]
+
{\rm c.c.}
\end{eqnarray}
Thus, using 
Eq.~(\ref{eq::F_NC0})
as a definition for
$g_{\rm nc}$,
we derive
\begin{eqnarray}
g_{\rm nc}
=
- \frac{\sqrt{2}}{\sqrt{3} |{\bf k}|^7}
\sum_{ {\bf R}^A }
	\cos ( {\bf p} \cdot {\bf R}^A )
	({\bf k} \cdot \nabla)^{4} 
	\left[
		V_{AB} \left( {\bf R}^A \right)
		+
		\Tilde V_{AB} \left( {\bf R}^A \right)
	\right].
\end{eqnarray} 
\end{widetext}
Assume additionally that the potential energies $V$ and 
$\Tilde V$
are short-range (that is, their characteristic space scale
$l_0$
is smaller than the 2D lattice constant). Then the expression for 
$g_{\rm nc}$
can be simplified 
\begin{eqnarray}
\label{eq::g3}
g_{\rm nc}
\approx
- \frac{\sqrt{2} {\cal Z}}{\sqrt{3} |{\bf k}|^7}
	\left[
	({\bf k} \!\cdot\! \nabla)^{4} \!
	\left(
		V_{AB}
		\!+\!
		\Tilde V_{AB}
	\right)
\right]_{{\bf R} = 0},
\end{eqnarray} 
where
${\cal Z}$
is the coordination number for a single site in the 2D lattice.

If $V$ and
$\Tilde V$
are characterized by an identical energy scale 
$\bar V$,
then
Eq.~(\ref{eq::g3})
can be used to write
$g_{\rm nc} \sim {\cal Z} \bar V/(|{\bf k}|^3 l_0^4)$.
This allows us to evaluate
$|c_6|
\sim
{\cal Z}^2 \bar V^2/(|{\bf k}|^6 l_0^8 \kappa_{{\bf p}})$,
where
$|c_6|$
is the
$m=3$
coefficient in
Eq.~(\ref{eq::F_NC}).

Let us conclude with the following observation. Our calculations in this
Appendix relied on the specific model of a layered solid. While the model
itself may be viewed as a very crude depiction of various layered
rare-earth tellurides, we believe that the discussed theoretical framework
can be extended beyond this realm. Instead of focusing of specific lattice
geometry, we would like to stress the essential premise embedded into the
formalism, namely the ability to split the lattice degrees of freedom
into two groups: those that host the CDW instability, and those that
``passively" react to the CDW formation. This grouping is quite obvious for
the layered structure shown in
Fig.~\ref{fig::model_lattice},
with its plane~A/B dichotomy. Perhaps less clear-cut in other situations,
yet it seems plausible that in a complicated multi-atomic elementary cell
different atoms display different degree of participation in the CDW
instability, providing an opportunity to introduce ``passive distortions"
of the lattice into a theoretical description. Such a reasoning can serve
as a broader justification for
Eq.~(\ref{eq::F_NC0}).

\section{Stiffness evaluation}
\label{app::stiffness}

In this Appendix we provide simple estimate for stiffness
$\kappa_{\bf p}$
that first emerges in
Eq.~(\ref{eq::F_NC0})
and then it is featured prominently both in
Sec.~\ref{sec::near_commensurate}
and
Appendix~\ref{app::NCCD}.
Within the model formulated in
Appendix~\ref{app::NCCD}
the stiffness is a characteristics of static deformations inside plane~B
lattice. The corresponding potential energy is
\begin{eqnarray}
F_B = 
{\cal N}_\perp
\sum_{{\bf R}, {\bf R}'}
	V_{BB} ({\bf R} - {\bf R}' + {\bf r} - {\bf r}'),
\end{eqnarray} 
where
${\bf R}$
and
${\bf R}'$
are lattice sites inside plane~B. Vector
${\bf r} = {\bf r} ({\bf R})$,
similar to 
${\bf r}^{A,B}$
in
Appendix~\ref{app::NCCD},
represents deviation of an atom from its position
${\bf R}$.
Likewise, vector
${\bf r}' = {\bf r} ({\bf R}')$
represents deviation of an atom from 
${\bf R}'$.

For weak periodic distortion
${\bf r} = 2 {\bf u} \cos ({\bf p}\cdot {\bf R} + \varphi)$,
we expand
$F_B$
in powers of
${\bf u}$.
Specifically, we write
$F_B = F^{(0)}_B + \delta F_B$,
where
\begin{eqnarray}
F^{(0)}_B
=
{\cal N}_\perp \sum_{{\bf R}, {\bf R}'} V_{BB} ({\bf R} - {\bf R}')
\end{eqnarray}
is the energy of unperturbed lattice, and
\begin{eqnarray}
\delta F_B =  F^{(2)}_B + O( | {\bf u} |^3),
\end{eqnarray} 
such that
$F_B^{(2)} = O( | {\bf u} |^2)$.
The first-order contribution 
$F_B^{(1)}$
vanishes. Indeed, one can write
\begin{eqnarray}
F_B^{(1)}
=
 2 {\cal N}_\perp \!\! \sum_{{\bf R}, {\bf R}'} \!\!
	\left[
		\cos ({\bf p}\cdot {\bf R} + \varphi)
		-
		\cos ({\bf p}\cdot {\bf R}' + \varphi)
	\right]
	\!\! \times
\\
\nonumber 
	({\bf u} \cdot \nabla) V_{BB} ({\bf R} - {\bf R}')
=
\\
\nonumber 
2 {\cal N}_\perp \!\! \sum_{{\bf R}, {\bf R}'} \!\!
	\left[
		\cos ({\bf p}\cdot ({\bf R} + {\bf R}') + \varphi)
		-
		\cos ({\bf p}\cdot {\bf R}' + \varphi)
	\right]
	\!\! \times
\\
\nonumber 
	({\bf u} \cdot \nabla) V_{BB} ({\bf R}).
\end{eqnarray} 
This expression vanishes 
due to
Eq.~(\ref{eq::qmomentum_conserv})
once summation over
${\bf R}'$
is performed.

To find 
$\kappa_{\bf p}$
we need to evaluate
\begin{eqnarray} 
F_B^{(2)}
=
8 {\cal N}_\perp \sum_{{\bf R}, {\bf R}'} 
\sin^2 ({\bf p}\cdot {\bf R} /2 + {\bf p} \cdot {\bf R}' + \varphi)
	\times
\\
\nonumber 
	\sin^2 ({\bf p}\cdot {\bf R}/2 )
	({\bf u} \cdot \nabla)^2 V_{BB} ({\bf R}).
\end{eqnarray} 
Since
$\sum_{{\bf R}'} 
\sin^2 ({\bf p}\cdot {\bf R} /2 + {\bf p} \cdot {\bf R}' + \varphi)
=
{\cal N}_{\rm 2D}/2$,
we derive
\begin{eqnarray} 
F_B^{(2)}
=
4 {\cal N}_{\rm 2D} {\cal N}_\perp
\sum_{{\bf R}}
	\sin^2 ({\bf p}\cdot {\bf R}/2 )
	({\bf u} \cdot \nabla)^2 V_{BB} ({\bf R}).
\end{eqnarray} 
Substituting
${\bf u} = |u| {\bf k} /|{\bf k}|$,
in agreement with
Eq.~(\ref{app::passive_u_def}),
we find the desired expression for the stiffness 
\begin{eqnarray}
\kappa_{\bf p} 
=
\frac{4}{|{\bf k}|^2}
\sum_{{\bf R}}
	\sin^2 ({\bf p}\cdot {\bf R}/2 )
	({\bf k} \cdot \nabla)^2 V_{BB} ({\bf R}).
\end{eqnarray}It depends on potential energy 
$V_{BB}$,
momentum
${\bf p}$,
and lattice symmetry. For us it is important that for small 
${\bf p}$
the stiffness vanishes. Note that corrections to
$\kappa_{\bf p}$
due to (weaker) inter-layer interaction were neglected in this calculation.
Yet for small
$\kappa_{\bf p}$
these contributions may be important.


\end{document}